\def \inparg{\leftskip = 40pt\rightskip = 40pt}
\def \outparg{\leftskip = 0 pt\rightskip = 0pt}

\def\npb{{Nucl.\ Phys.\ }{\bf B}}
\def\plb{{Phys.\ Lett.\ }{ \bf B}}

\def\prd{{Phys.\ Rev.\ }{\bf D}}

\def\prl{Phys.\ Rev.\ Lett.\ }

\def\frak#1#2{{\textstyle{{#1}\over{#2}}}}

\def\epsilonbar{\bar\epsilon}
\def\lambdabar{\bar\lambda}
\def\sigmabar{\bar\sigma}
\def\phibar{\bar\phi}
\def\psibar{\bar\psi}
\def\Ncal{{\cal N}}
\def\Fbar{\bar F}

\def\sy{supersymmetry}
\def\sic{supersymmetric}
\def\pa{\partial}  
\input harvmac
\input epsf
{\nopagenumbers
\line{\hfil LTH 666}
\line{\hfil hep-th/0509089}
\vskip .5in
\centerline{\titlefont Renormalisation of supersymmetric gauge theory}
\centerline{\titlefont in the uneliminated component formalism}
\vskip 1in
\centerline{\bf I.~Jack, D.R.T.~Jones and L.A.~Worthy}
\bigskip
\centerline{\it Department of Mathematical Sciences,
University of Liverpool, Liverpool L69 3BX, U.K.}
\vskip .3in
We show that the renormalisation of the ${\cal N}=1$ supersymmetric gauge 
theory when working in the component formalism, without eliminating 
auxiliary fields and using a standard covariant gauge, requires a non-linear
renormalisation of the auxiliary fields. 
\Date{September 2005}}

\newsec{Introduction}

The renormalisation of ${\cal N}=1$ supersymmetric gauge theory is 
certainly well-understood in the superfield formalism both in terms of formal 
analysis (for example 
Ref.~\ref\piguet{O.~Piguet, hep-th/9611003})
and 
practical calculations  (for example
Ref.~\ref\jjn{I. Jack, D.R.T. Jones and C.G. North, \npb 486 (1997) 479}). 
In accordance with the 
non-renormalisation theorem the superpotential is 
unrenormalised, leading to the standard expression for the Yukawa-coupling
$\beta$-function in terms of the chiral superfield anomalous dimension.  
However, a feature of the superfield formalism which is often overlooked is the
necessity for a non-linear renormalisation of the vector superfield
\ref\storey{J.W.~Juer and D.~Storey,  
\plb119 (1982) 125; \npb216 (1983) 185}. 

In fact, as we shall see, the renormalisation program is perhaps most 
straightforwardly implemented in terms of component fields and in 
the case where the auxiliary fields $F$ and $D$ are
eliminated using their equations of motion. It is well-documented in this
case that the Lagrangian is multiplicatively renormalisable. From a practical 
point of view, moreover,   
although a softly-broken \sic\  theory {\it can\/}
be treated using superfields via spurion techniques,  calculations in
the MSSM are generally carried out using the eliminated component
formalism. Now since the elimination of $F$ and $D$
gives rise to non-linear  terms in the supersymmetry
transformations of the physical fields, one might expect that  the
renormalisation program would be at least as simple in terms of the uneliminated
formalism.  Indeed, the uneliminated formalism has been employed for
effective potential  
calculations~\ref\miller{R.D.C.~Miller, \npb 229 (1983) 189} 
and in calculations of
the $\beta$-function for soft  $(\hbox{mass})^2$ 
terms~\ref\jjp{I. Jack, D.R.T. Jones and S. Parsons, \prd 62 (2000) 125022, 
{\it ibid}  {\bf D}63 (2001) 075010}.  Our purpose here
is simply to show how the uneliminated  formalism requires some care
in that (in a conventional covariant gauge) the theory is once again not 
multiplicatively renormalisable in the conventional sense; additional 
counter-terms are required which do not correspond to terms in the
original Lagrangian but which can be generated by non-linear field 
renormalisations. However, these non-linear field renormalisations
appear to be distinct from those of Ref.~\storey, since they only appear
in the presence of chiral matter whereas the latter arise even in the pure
gauge case.
  
We also consider what happens in the light-cone gauge, which is, in a sense 
we shall explain, ``more supersymmetric'' than the conventional covariant 
gauge~\ref\dctj{D.M. Capper and D.R.T. Jones, \prd 31 (1985) 3295}.    

\newsec{Renormalisation}

The Lagrangian is given in components by 
\eqn\lagrana{\eqalign{
 S_{\hbox{unel}}
=&\int d^4x\Bigl[-\frak14F^{\mu\nu A}F^A_{\mu\nu}-i\lambdabar^A\sigmabar^{\mu}
(D_{\mu}\lambda^A)+\frak{1}{2}(D^A)^2\cr
&+F^i F_i -i\psibar\sigmabar^{\mu}D_{\mu}\psi-D^{\mu}\phibar D_{\mu}\phi
+g\phibar R^A\phi D^A +i\sqrt2g(\phibar \lambda\psi-\psibar\lambdabar\phi)\cr
&+F^iW_i+F_iW^i-\frak12W_{ij}\psi^i\psi^j-\frak12W^{ij}\psi_i\psi_j\Bigr],\cr}}
where 
\eqn\Wdef{
W(\phi)=\frak{1}{6}Y^{ijk}\phi_i\phi_j\phi_k}
is the superpotential, assumed cubic in $\phi$ for
renormalisability, $W_i\equiv {\pa W\over{\pa \phi^i}}$, and
the lowering of indices indicates complex conjugation, so that
$W^i=(W_i)^*$. For simplicity we omit possible linear and quadratic terms. 
The chiral fields transform according to a representation
$R$ of the gauge group and we write $\lambda=\lambda^AR^A$.  
If we eliminate the auxiliary fields $F$ and $D$ using their equations of 
motion:
\eqn\eqmotion{\eqalign{
D^A + g\phibar R^A\phi &=0,\cr
F^i + W^i &= 0,\cr}}
we obtain
the eliminated Lagrangian, given in components by
\eqn\lagranb{\eqalign{
 S_{\hbox{el}}=&\int d^4x\Bigl[
-\frak14F^{\mu\nu A}F_{\mu\nu}^A-i\lambdabar^A\sigmabar^{\mu}
(D_{\mu}\lambda^A)\cr
&-i\psibar\sigmabar^{\mu}D_{\mu}\psi-D^{\mu}\phibar D_{\mu}\phi
-\frak12g^2(\phibar R^A\phi)(\phibar R^A\phi)
 +i\sqrt2g(\phibar \lambda\psi-\psibar\lambdabar\phi)\cr
&-W^iW_i-\frak12W_{ij}\psi^i\psi^j-\frak12W^{ij}\psi_i\psi_j\Bigr].\cr}}
In either case we use the standard gauge-fixing term
\eqn\gafix{
S_{\rm{gf}}={1\over{2\alpha}}\int d^4x (\pa.A)^2}
with its associated
ghost terms.
The theory in the eliminated case is rendered finite by replacing fields
and couplings in Eq.~\lagranb\ by their corresponding bare versions. We have  
\eqn\bare{\eqalign{ \lambda^A_B=Z_{\lambda}^{\frak12}\lambda^A,
\quad
A^{A}_{\mu B}=Z_A^{\frak12}A^{A}_{\mu},\quad \phi_B=Z_{\phi}^{\frak12}\phi,
\quad \psi_B=&Z_{\psi}^{\frak12}\psi,\cr
\quad g_B=Z_gg,
\quad Y_B^{ijk}=\left(Z_{\Phi}^{-\frak12}\right)^i_l
\left(Z_{\Phi}^{-\frak12}\right)^j_m
\left(Z_{\Phi}^{-\frak12}\right)^k_nY^{lmn}.\cr}}
Here $Z_{\Phi}$ is the 
renormalisation constant for the chiral superfield $\Phi$ so that the 
result for $Y_B$ is the consequence of the non-renormalisation theorem.
In general, however,  when working in a standard covariant gauge in components, 
$Z_{\phi,\psi,\Phi}$ are all different; at one loop, in fact, we have
\eqn\Zgg{\eqalign{
Z_{\lambda}&=1-2g^2L[\alpha C(G) + T(R)],\cr
Z_A&=1+g^2L[(3-\alpha)C(G)-2T(R)],\cr
Z_g&=1+g^2L\left[T(R)-3C(G)\right],\cr
Z_{\phi}&=1+L\left[-Y^2+2(1-\alpha)g^2C(R)\right],\cr
Z_{\psi}&=1+L\left[-Y^2-2(1+\alpha)g^2C(R)\right],\cr
Z_{\Phi}&=1+L\left[-Y^2+4g^2C(R)\right],\cr}}
where 
\eqn\groupdefs{\eqalign{
\left(Y^2\right)^i{}_j=&Y^{ikl}Y_{jkl}\cr
C(R)=R^AR^A,&\quad T(R)\delta^{AB}=\Tr[R^AR^B],\cr}}
$C(G)$ is the  adjoint Casimir and 
(using dimensional regularisation with $d=4-\epsilon$)
$L={1\over{16\pi^2\epsilon}}$. 
But now what happens if we work with the uneliminated form of the action? 
We might expect the theory to be rendered finite by replacing fields and 
couplings in Eq.~\lagrana\ by corresponding bare versions (now we also 
need $F_B=(Z_F)^{\frak12}F$, $D_B=(Z_D)^{\frak12}D$ of course). It 
is not difficult to see, however, 
that there are one-loop diagrams with 2 $\phi$ and 
2 $\phibar$ external fields for which there are no counter-term diagrams
in this case (while in the eliminated case, counterterms are supplied by the 
$W^iW_i$ term). We also find that the $F\phi^2$ and $D\phibar\phi$ 
terms are not rendered finite by the renormalisation constants given above.
\bigskip
\epsfysize= 1in
\centerline{\epsfbox{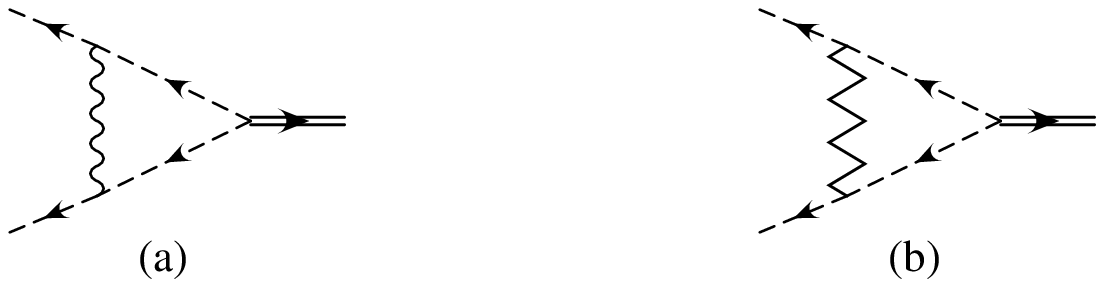}}
\inparg
{\it \noindent Fig.~1: Diagrams with one $F$, two scalar lines. Dashed, full,
double full, wavy, full/wavy, zigzag lines represent $\phi$, $\psi$, $F$,
$A$, $\lambda$, $D$ propagators respectively.}
\medskip
\outparg

To be precise, the results for the graphs in Fig.~1 are:
\eqna\figone
$$
\eqalignno{
\Gamma^1_a=&-\frak12\alpha Lg^2 \left(Y_{ijk}[C(R)F]^i\phi^j\phi^k
-2Y_{ijk}F^i[C(R)\phi]^j\phi^k\right), &\figone a\cr
\Gamma^1_b=&-\frak12Lg^2 \left(Y_{ijk}[C(R)F]^i\phi^j\phi^k
-2Y_{ijk}F^i[C(R)\phi]^j\phi^k\right), &\figone b\cr}
$$
\bigskip
\epsfysize= 1in   
\centerline{\epsfbox{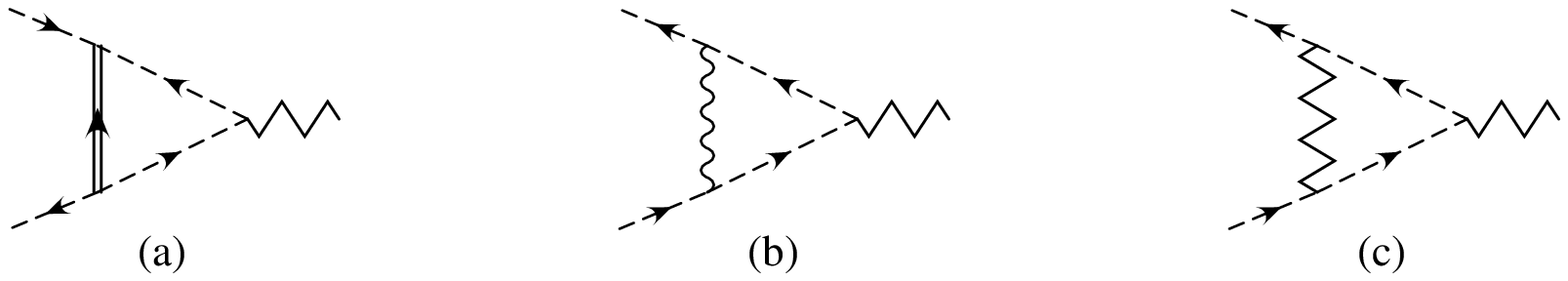}}
\inparg
{\it \noindent Fig.~2: Diagrams with one $D$, two scalar lines.}
\medskip
\outparg

and the results for the graphs in Fig.~2 are
\eqn\figtwo{\eqalign{
\Gamma^2_a=&L \phibar Y^2D\phi,\cr
\Gamma^2_b=&2\alpha Lg^2\phibar[C(R)-\frak12C(G)]D\phi,\cr
\Gamma^2_c=&-2Lg^2\phibar[C(R)-\frak12C(G)]D\phi,\cr}}
where $D=D^AR^A$.
\medskip
\epsfysize= 1.75in
\centerline{\epsfbox{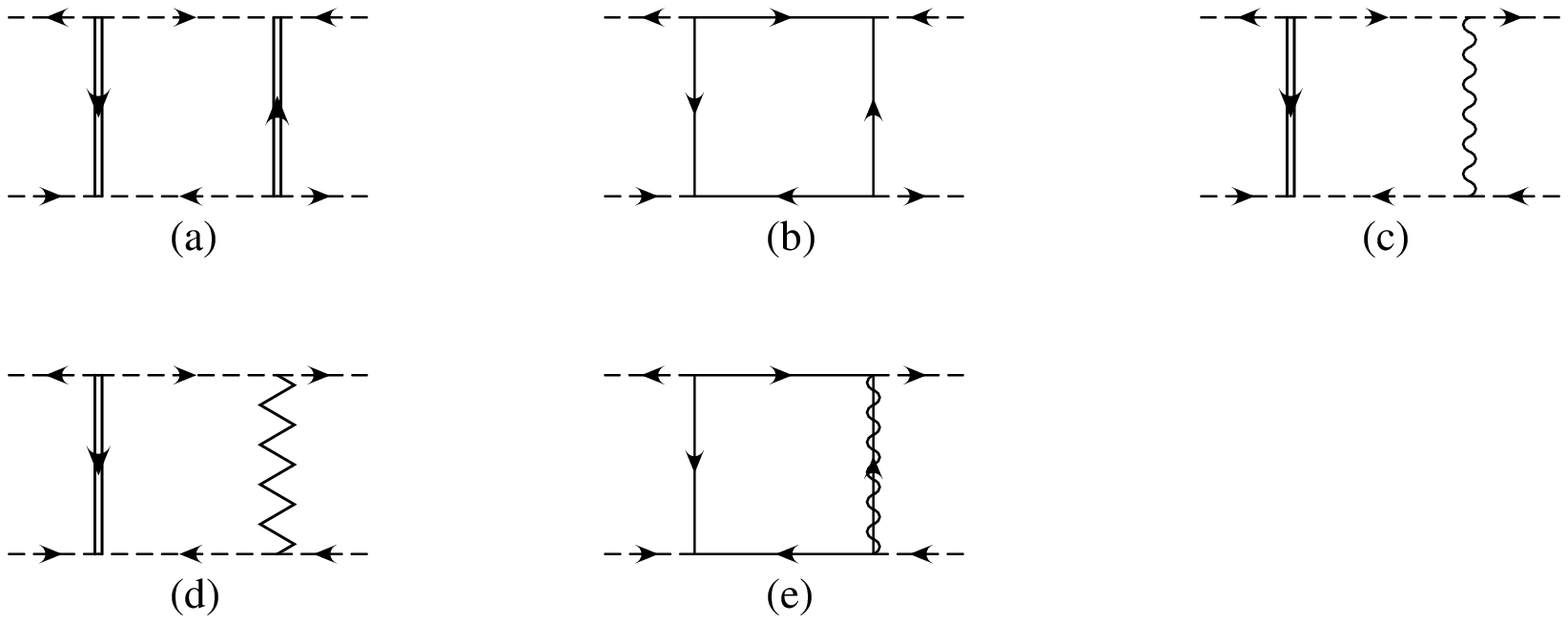}}
\inparg
{\it \noindent Fig.~3: Diagrams with 2 $\phi$, 2 $\phibar$ lines
and 2 or 4 Yukawa vertices.}
\medskip
\outparg
The results for the graphs in Fig.~3 are
\eqn\figthree{\eqalign{ 
\Gamma^3_a=&LY_{imn}Y_{jpq}Y^{kmp}Y^{lnq}\phi^i\phi^j\phi_k\phi_l,\cr
\Gamma^3_b=&-LY_{imn}Y_{jpq}Y^{kmp}Y^{lnq}\phi^i\phi^j\phi_k\phi_l,\cr
\Gamma^3_c=&-\frak12\alpha Lg^2Y_{ijm}Y^{kln}[C(R)]^m{}_n
\phi^i\phi^j\phi_k\phi_l,\cr
\Gamma^3_d=&\frak12 Lg^2Y_{ijm}Y^{kln}[C(R)]^m{}_n\phi^i\phi^j\phi_k\phi_l,\cr
\Gamma^3_e=&-2 Lg^2Y_{ijm}Y^{kln}[C(R)]^m{}_n\phi^i\phi^j\phi_k\phi_l.\cr}}
\medskip
\epsfysize= 3in
\centerline{\epsfbox{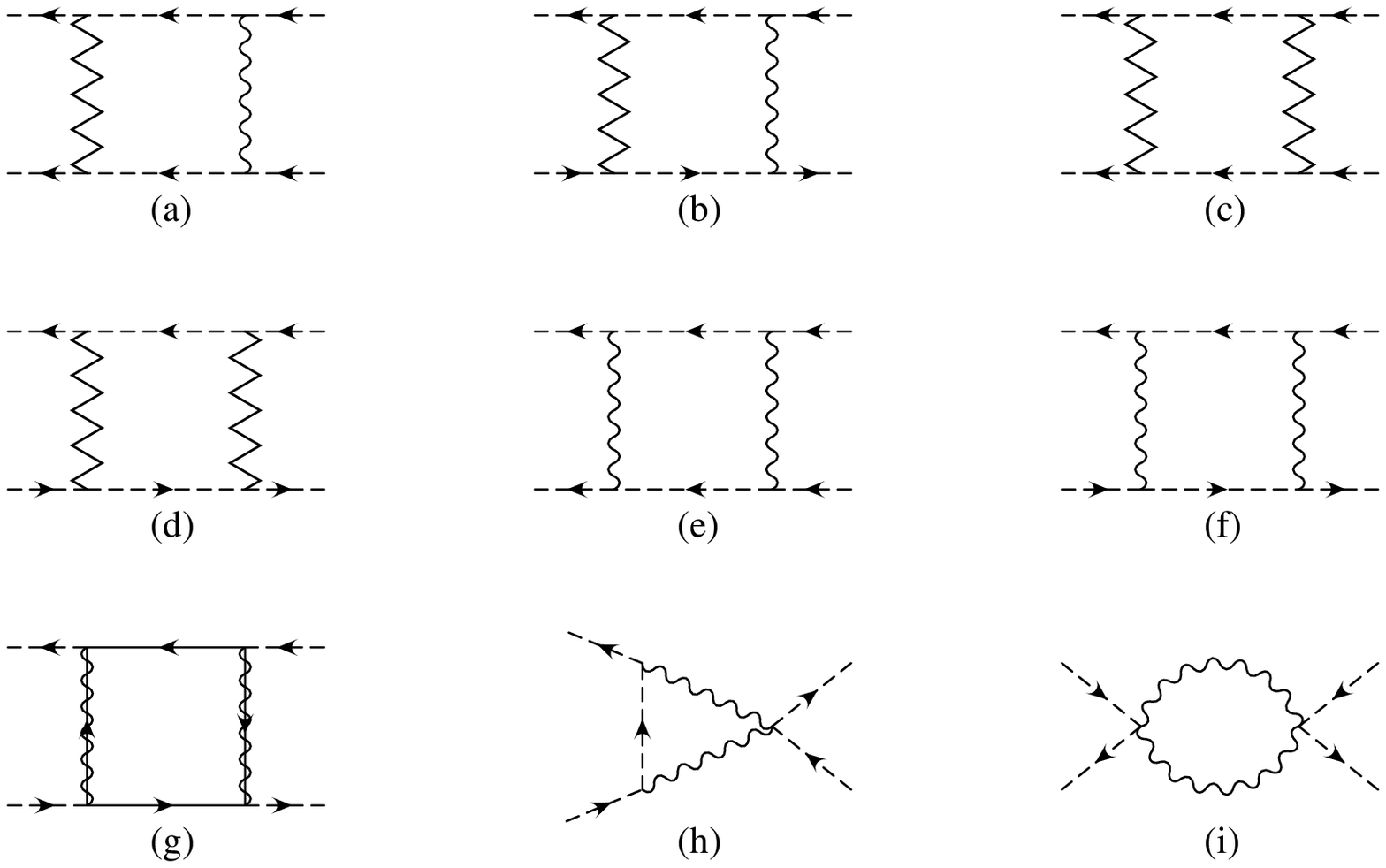}}
\inparg
{\it \noindent Fig.~4: Diagrams with 2 $\phi$, 2 $\phibar$ lines
and 4 gauge vertices.}
\medskip
\outparg
The results for the graphs in Fig.~4 are
\eqn\figfour{\eqalign{
\Gamma^4_a=&2\alpha g^4L\left(\phibar R^AR^B\phi\right) \left(\phibar R^AR^B\phi\right),\cr
\Gamma^4_b=&-2\alpha g^4L\left(\phibar R^AR^B\phi\right) \left(\phibar R^BR^A\phi\right),\cr
\Gamma^4_c=&g^4L\left(\phibar R^AR^B\phi\right) \left(\phibar R^AR^B\phi\right),\cr
\Gamma^4_d=&g^4L\left(\phibar R^AR^B\phi\right) \left(\phibar R^BR^A\phi\right),\cr
\Gamma^4_e=&\alpha^2 g^4L\left(\phibar R^AR^B\phi\right) \left(\phibar R^AR^B\phi\right),\cr
\Gamma^4_f=&\alpha^2 g^4L\left(\phibar R^AR^B\phi\right) \left(\phibar R^BR^A\phi\right),\cr
\Gamma^4_g=&-8g^4L\left(\phibar R^AR^B\phi\right) \left(\phibar R^BR^A\phi\right),\cr
\Gamma^4_h=&-2\alpha^2 g^4L\left(\phibar R^AR^B\phi\right) 
\left(\phibar R^AR^B\phi+\phibar R^BR^A\phi\right),\cr
\Gamma^4_i=&(3+\alpha^2) g^4L\left(\phibar R^AR^B\phi\right) 
\left(\phibar R^AR^B\phi+\phibar R^BR^A\phi\right).\cr}}
The results for the graphs contributing to the remaining interaction terms in 
Eq.~\lagrana\ are the same as in the eliminated case so we shall not give 
detailed results. The renormalisation constants $Z_{\phi,\psi,A}$ are also the
same as in the eliminated case, and in addition we have
\eqn\ZF{
Z_F=1-LY^2, \qquad Z_D=1-2Lg^2T(R).}
We find that 
\eqn\leftover{\eqalign{
&\left[\Gamma_1
+\frak12Y_{Bijk}F_B^i\phi_B^j\phi_B^k + \hbox{(c.c.)}\right]
+\Gamma_2+\Gamma_3+\Gamma_4
+g_B\phibar_BD_B\phi_B\cr
&=\Bigl\{\frak12Y_{ijk}F^i\phi^j\phi^k
-\frak12(\alpha+3)g^2L\Bigl[Y_{ijk}[C(R)F]^i\phi^j\phi^k
\cr
&+\frak12
Y_{ijm}Y^{kln}[C(R)]^m{}_n\phi^i\phi^j\phi_k\phi_l\Bigr]+(\hbox{c.c.})\Bigr\}\cr
&+g\phibar D\phi
-(\alpha+2)C(G)g^3L\left[\phibar D\phi+g\left(\phibar R^A\phi\right)
\left(\phibar R^A\phi\right)\right].\cr}}
The residual divergence cancels if we substitute the 
equations of motion, Eq.~\eqmotion,   
for $D^A$ and $F^i$, as we would expect.

Alternatively, it is clear that these remaining divergences 
can all be cancelled by making the
nonlinear renormalisations
\eqn\redefs{\eqalign{
(F_B)_i=&(Z_F^{\frak12}
F)_i+\frak12(\alpha+3)g^2L[C(R)]^l{}_iY_{ljk}\phi^j\phi^k,\cr
(D_B)^A=&(Z_D^{\frak12}D)^A+(\alpha+2)C(G)g^3L\phibar R^A\phi.\cr}}

A similar phenomenon was observed in a study of the renormalisation 
of ${\cal N} = \frak{1}{2}$ theories, presented in 
Refs.~\ref\jjw{I.~Jack, D.R.T.~Jones
and L.A.~Worthy, \plb 611 (2005) 199; \prd72 (2005) 065002}; though in the case 
without a superpotential considered there, application of the equation of 
motion for $F$ is rather trivial, since 
the equation of motion for $\Fbar$ gives 
$F=0$. In the ${\cal N} = \frak{1}{2}$ case, however,  
a further field redefinition (of the gaugino field $\lambda$) is necessary, 
and this redefinition has no analogy in the ${\cal N} = 1$ case 
considered here.

\newsec{The light-cone gauge}

It is interesting to reconsider the above calculations in the light-cone gauge, 
corresponding to the $\alpha\to 0$ limit of
\eqn\lcgf{S_{\rm{gf}}={1\over{2\alpha}}\int d^4x (n.A)^2, 
\quad\hbox{with}\quad n^2 = 0.
}
In the light-cone gauge one again has a choice between an eliminated 
and an uneliminated formalism, distinct from that associated with 
the auxiliary fields of \sy. Choosing $n = n^{-}$, the light-cone gauge 
corresponds to $A^{+}=0$ and  the field $A^{-}$ is non-propagating 
and can be eliminated by its equation of motion.   Moreover, the 
condition $A^{+}=0$ is preserved by the subset of \sy\ transformations 
corresponding to setting the infinitesimal spinor $\epsilon$ governing these 
transformations to be $\epsilon = \epsilon^{+}$. 
(This is reminiscent of 
${\cal N} =\frak{1}{2}$ \sy~\ref\seib
{H. Ooguri and C. Vafa, Adv. Theor. Math. Phys, 7 (2003) 53\semi
J. de Boer, P.A. Grassi and P. van Nieuwenhuizen, 
\plb 574 (2003) 98\semi
N.~Seiberg, JHEP{\bf 0306} (2003) 010}, where the 
action is invariant under \sy\ transformations with 
respect to $\epsilon$, but with $\epsilonbar = 0$). 
As a consequence, 
one finds in the light-cone gauge that
\eqn\Zgg{\eqalign{Z_{\lambda}&= 1-2g^2L\left[T(R)-3C(G)\right],\cr
Z_{\phi}&= Z_{\psi} = 1+L\left[-Y^2+4g^2C(R)\right],\cr}}
reflecting the preservation of (half the) \sy\ by the gauge.  

Light-cone gauge QCD was discussed in 
Ref.~\ref\leem{H.C.~Lee and M.S.~Milgram, \prl 55 (1985) 2122},where 
it was shown that 
a computation of  the gauge two-point function in the $A^{-}$-uneliminated 
formalism leads to  
divergent structures not corresponding to terms in the Lagrangian, which 
however vanish if the equation of motion for $A^{-}$ is applied. 
So this is completely analogous to the situation we found above.

Returning to the \sic\ theory, 
we  have recalculated Eq.~\figone{}\ in the uneliminated light-cone gauge; 
Eq.~\figone{b}\ is manifestly unchanged but $\Gamma^1_a = -\Gamma^1_b$  
so that there is no 1PI divergence, as in the superfield case.
$Z_{\phi}$ now corresponds to the \sic\ result (as indicated above in Eq.~3.2), 
but $Z_F$ remains the same as in 
the covariant gauge case and so we obtain 

\eqn\lcgams{
\Gamma_1 +\frak12Y_{Bijk}F_B^i\phi_B^j\phi_B^k+(\hbox{c.c}) = 
\frak12Y_{ijk}F^i\phi^j\phi^k-g^2LY_{ijk}[C(R)F]^i\phi^j\phi^k+(\hbox{c.c}),
} 
or more generally (instead of Eq.~\leftover)

\eqn\leftover{\eqalign{
&\left[\Gamma_1+\frak12Y_{Bijk}F_B^i\phi_B^j\phi_B^k+(\hbox{c.c.})\right]
+\Gamma_2+\Gamma_3+\Gamma_4
+g_B\phibar_BD_B\phi_B\cr
=&\Bigl\{\frak12Y_{ijk}F^i\phi^j\phi^k
-g^2L\Bigl[Y_{ijk}[C(R)F]^i\phi^j\phi^k\cr
&
+\frak12Y_{ijm}Y^{kln}[C(R)]^m{}_n\phi^i\phi^j\phi_k\phi_l\Bigr]
+(\hbox{c.c})\Bigr\}\cr
&+g\phibar D\phi-C(G)g^3L\left[\phibar D\phi+g\left(\phibar R^A\phi\right)
\left(\phibar R^A\phi\right)\right].\cr}}
Once again the residual divergence vanishes upon application of the equations 
of motion for $F$, $D$, or via a non-linear renormalisation 
corresponding to setting $\alpha = -1$ in Eq.~\redefs.

\newsec{Conclusions}

We have seen that for ${\cal N} = 1$ theories  the renormalisation
program, when carried out in the $F,D$  uneliminated formalism, contains
some subtlety in that  divergent terms of a form not present in the
original Lagrangian  are generated. These terms can, in fact, be
eliminated  either by means of non-linear field redefinitions (or
renormalisations)  or by imposing the equations of motion for $F,D$. We
also recalled how  an analogous phenomenon occurs in the light-cone
gauge, where  the  r\^ole of the non-propagating $F,D$ fields is played
by the $A^-$ gauge field. We believe that there is some pedagogical
justification for clarifying these somewhat subtle features of the uneliminated 
form of the familiar $\Ncal=1$ supersymmetric theory. Moreover,
this renders unsurprising the non-linear redefinition of $\Fbar$ found 
necessary in the $\Ncal=\frak12$ case\jjw. In particular, it is interesting 
that in both cases the non-linear redefinition is gauge-parameter 
dependent. The phenomenon may also help to elucidate the 
additional redefinition of $\lambda$ found to be required in the 
$\Ncal=\frak12$ case.

\listrefs
\bye